\documentclass[twocolumn,showpacs,prb,amsfonts,amsmath,amssymb,floatfix,groupedaddress]{revtex4}
\usepackage{color}
\usepackage{hhline}
\usepackage{mathrsfs}
\usepackage{graphicx}
\usepackage{dcolumn}
\usepackage{bm}
\usepackage{multirow}
\usepackage{booktabs}
\usepackage{afterpage}
\usepackage{amsmath}

\begin{document}

\title{First-principles study of the pressure and crystal-structure dependences of the superconducting
transition temperature in compressed sulfur hydrides}

\author{Ryosuke Akashi$^{1}$}
\thanks{Corresponding author}
\author{Mitsuaki Kawamura$^{1}$}
\author{Shinji Tsuneyuki$^{1,2}$}
\affiliation{$^1$Department of Physics, The University of Tokyo, Hongo, Bunkyo-ku, Tokyo 113-0033, Japan}
\affiliation{$^2$Institute of Solid State Physics, The University of Tokyo, Kashiwa, Chiba 277-8581, Japan}
\author{Yusuke Nomura$^{3,4}$}
\author{Ryotaro Arita$^{3,5}$}
\affiliation{$^3$RIKEN Center for Emergent Matter Science, Wako, Saitama 351-0198, Japan}
\affiliation{$^4$Department of Applied Physics, The University of Tokyo, Hongo, Bunkyo-ku, Tokyo 113-8656, Japan}
\affiliation{$^5$JST ERATO Isobe Degenerate $\pi$-Integration Project, Advanced Institute for Materials Research (AIMR), Tohoku University, Sendai, Miyagi 980-8577, Japan}

\date{\today}
\begin{abstract}
We calculate superconducting transition temperatures ($T_{\rm c}$) in sulfur hydrides H$_{2}$S and  H$_{3}$S from first principles using the density functional theory for superconductors. At pressures of $\lesssim$~150~GPa, the high values of $T_{\rm c}$ ($\geq$130~K) observed in the recent experiment [A. P. Drozdov, M. I. Eremets, and I. A. Troyan, arXiv:1412.0460] are accurately reproduced by assuming that H$_{2}$S decomposes into $R3m$-H$_{3}$S and S. For the higher pressures, the calculated $T_{\rm c}$s for $Im\overline{3}m$-H$_{3}$S are systematically higher than those for $R3m$-H$_{3}$S and the experimentally observed maximum value (190~K), which suggests the possibility of another higher-$T_{\rm c}$ phase. We also quantify the isotope effect from first principles and demonstrate that the isotope effect coefficient can be larger than the conventional value (0.5) when multiple structural phases energetically compete.
\end{abstract}
\pacs{74.62.Fj, 74.20.-z, 74.25.Kc, 74.70.-b}

\maketitle
\section{Introduction}
\label{sec:intro}
Investigating compounds containing light elements has been a simple and powerful guiding principle for discovery of high-temperature superconductors. According to the BCS theory,~\cite{BCS} the superconducting transition temperature ($T_{\rm c}$) is scaled by the phonon frequency and therefore light atoms are advantageous for achieving high $T_{\rm c}$. Despite its simplicity, this principle has been surprisingly successful as represented by the discoveries of superconductivity in doped fullerene solids,~\cite{fullerene} magnesium diboride,~\cite{MgB2} lithium under pressure~\cite{Li-pressure-Shimizu, Li-pressure-Struzhkin} and boron-doped diamond.~\cite{B-diamond-Ekimov,B-diamond-Takano} Along this principle, possible superconductivity in compressed hydrogen and hydrogen compounds has been explored as an extreme case.~\cite{Ashcroft-H,Ashcroft-hydrides,Feng-Ashcroft-SiH4,Tse-SiH4-SC,Eremets-Tse-SiH4-exp2008,Degtyareva-PtH-exp2009,Li-Gao-SiH4H22-PNAS2010,Jin-Ma-Cui-Si2H6-PNAS2010,Jose-Marques-Goedecker-Si2H6-PRL2012,Cudazzo-hydrogen,Li-Ma-H2S-struct,Duan-H3S-struct, Scheler-Degtyareva-PtH-PRB2011, Zhou-Oganov-PtH-PRB2011, Kim-Pickard-Needs-PtH-PRL2011, Tse-SnH4-PRL2007, Gao-Oganov-SnH4-PNAS2010,Goncharenko-Erements-Tse-AlH3-PRL2008,Gao-Oganov-GeH4-PRL2008,Szczesniak-GeH4-SSComm2013,Gao-Bergara-GaH3-PRB2011,Szczesniak-GaH3-SSTech2014,Tse-CaH-PNAS,Lonie-Zurek-MgH-PRB2013, Wang-Yao-Ma-BeH2-JChemPhys2014,Zhou-Ma-Cui-KH6-PRB2012,Kim-Ahuja-trend-PNAS2010,Gao-Ashcroft-Bergara-NbH-PRB2013}

Recently, it has been discovered that H$_{2}$S exhibits superconductivity under high pressures at 190K (Ref.~\onlinecite{Eremets}). Being the new record of the superconducting transition temperature ($T_{\rm c}$), this report has immediately aroused intense debate.\cite{Durajski-Li-preprint,Hirsch-Marsiglio-preprint,Mazin-preprint, Duan-preprint,Pickett-preprint} Several facts imply that this superconducting phase is induced by the conventional mechanism due to the vibrations of hydrogen atoms:  The observed $T_{\rm c}$ is subject to the hydrogen isotope effect~\cite{Eremets}; in prior to the experimental discovery, there was an {\it ab initio} calculation which predicted strong electron-phonon coupling~\cite{Li-Ma-H2S-struct}; the electronic bandwidth is so large that the Migdal approximation seems valid.~\cite{Migdal} However, some puzzling results have also been exposed. First, the crystal structure realized in the experimental situation has not been specified. If we estimate $T_{\rm c}$ of H$_{2}$S using the conventional McMillan formula \cite{Li-Ma-H2S-struct,Allen-Dynes} with the empirical Coulomb parameter $\mu^{\ast}$=0.13, the calculated value is too low compared with the experimentally observed value. It has also been proposed that H$_{3}$S phase instead emerges under high pressures,\cite{Mazin-preprint, Pickett-preprint} where the electron-phonon coupling is thought to be stronger than in H$_{2}$S.~\cite{Duan-H3S-struct} Second, anomalously large hydrogen isotope effect coefficient $\alpha$$\sim$1.0 has been observed. Although it has been hypothesized that the unharmonic effect on the lattice dynamics has some role~\cite{Pickett-preprint} or that different structures emerge in H$_{2}$S and D$_{2}$S (sulfur di-deuteride),~\cite{Hirsch-Marsiglio-preprint} this anomaly remains an open question.

To further investigate the above points, we need to address not only the electron-phonon interaction but also the electron-electron Coulomb interaction in the H$_{x}$S systems. Accurate evaluation of the impact of the pair-breaking Coulomb repulsion is vital because this governs the absolute value of $T_{\rm c}$, as well as $\alpha$.~\cite{Morel-Anderson, Garland-isotope, Carbotte-review} In addition, experimentally realized pressure range is rather out of that in the previous {\it ab initio} studies and therefore more thorough investigations of the pressure dependence of the superconducting properties are desired.

In this Article, we present an {\it ab initio} study on the superconductivity in solid H$_{2}$S and H$_{3}$S covering the experimental pressure range. In the standard Migdal-Eliashberg theory,~\cite{Migdal,ME} the effect of the electron-electron Coulomb interaction is practically treated with an empirical parameter $\mu^{\ast}$. To incorporate this effect non-empirically, we utilize the density functional theory for superconductors (SCDFT~\cite{SCDFTI,SCDFTII}). With this theory, we can calculate $T_{\rm c}$ and $\alpha$ without any empirical parameter, which can be directly compared with the experimental data. 

\section{Method}
\label{sec:method}
To calculate $T_{\rm c}$ from first principles, we employed the SCDFT gap equation given by
\begin{eqnarray}
\Delta_{n{\bf k}}\!=\!-\mathcal{Z}_{n\!{\bf k}}\!\Delta_{n\!{\bf k}}
\!-\!\frac{1}{2}\!\sum_{n'\!{\bf k'}}\!\mathcal{K}_{n\!{\bf k}\!n'{\bf k}'}
\!\frac{\mathrm{tanh}[(\!\beta/2\!)\!E_{n'{\bf k'}}\!]}{E_{n'{\bf k'}}}\!\Delta_{n'\!{\bf k'}}.
\label{eq:gap-eq}
\end{eqnarray}
Here, $n$ and ${\bf k}$ denote the band index and crystal momentum, respectively, $\Delta_{n{\bf k}}$ is the gap function, and $\beta$ is the inverse temperature. The energy $E_{n {\bf k}}$ is defined as $E_{n {\bf k}}$=$\sqrt{\xi_{n {\bf k}}^{2}+\Delta_{n {\bf k}}^{2}}$ and $\xi_{n {\bf k}}$ is the one-electron energy with respect to the Fermi level calculated with the normal Kohn-Sham equation. The functions $\mathcal{Z}$ and $\mathcal{K}$ are called exchange-correlation kernels, which describe the effects of the interactions. The nondiagonal kernel $\mathcal{K}$ consists of two parts $\mathcal{K}$$=$$\mathcal{K}^{\rm ph}$$+$$\mathcal{K}^{\rm el}$ representing the electron-phonon and electron-electron interactions, respectively. The diagonal kernel $\mathcal{Z}$$=$$\mathcal{Z}^{\rm ph}$ represents the mass renormalization of the normal-state band structure due to the phonon exchange. Using these kernels,~\cite{SCDFTI, SCDFTII} the conventional strong-coupling superconductivity can be treated with the level of the Migdal-Eliashberg theory.~\cite{Migdal,ME} In particular, the electronic nondiagonal kernel $\mathcal{K}^{\rm el}$ describes the screened electron-electron Coulomb interaction, where the dynamical screening effects are incorporated within the random-phase approximation.~\cite{Akashi-plasmon-PRL, Akashi-plasmon-JPSJ} We can therefore evaluate effects of the static Coulomb repulsion suppressing the pairing, as well as the plasmon superconducting mechanism.~\cite{Takada-plasmon-JPSJ}
\begin{table}[b]
\caption[t]
{Pressure settings and the corresponding input structures for the calculations. We observed that it is difficult to achieve the numerical convergence in the phonon calculations for the calculations for H$_{3}$S at 190~GPa since it is near the second-order structural transition point.~\cite{Duan-H3S-struct}}
\begin{center}
\label{tab:struct}
\tabcolsep = 1mm
\begin{tabular}{|l |c|c|c|c|c|c|c|} \hline
 $P$ [GPa]&130&150&170&190&210&230&250 \\ \hline
H$_{2}$S & \multicolumn{2}{c|}{$P1$~\cite{Li-Ma-H2S-struct}} & \multicolumn{5}{c|}{$Cmca$~\cite{Li-Ma-H2S-struct}} \\ \hline
H$_{3}$S & \multicolumn{3}{c|}{$R3m$~\cite{Duan-H3S-struct}} &. . . &\multicolumn{3}{c|}{$Im\overline{3}m$~\cite{Duan-H3S-struct}} \\
 \hline
\end{tabular}
\end{center}
\end{table}

We calculated the electronic states, phonon frequencies, electron-phonon and electron-electron interactions and $T_{\rm c}$ for H$_{2}$S and H$_{3}$S at various pressures. Our calculations were performed with the generalized-gradient approximation using the exchange-correlation potential with the Perdew-Burke-Ernzerhof parametrization.~\cite{GGAPBE} We used {\it ab initio} plane-wave pseudopotential calculation codes {\sc QUANTUM ESPRESSO}~\cite{QE} for the electroic structure, dynamical matrix and electron-phonon coupling. The input crystal structures at respective pressures were the optimum ones predicted in the previous {\it ab initio} calculations, which are summarized in Table \ref{tab:struct}. For all the conditions, we optimized the atomic configurations and cell parameters with respect to enthalpy under fixed pressures. Phonon frequencies and electron-phonon interactions were calculated based on the density functional perturbation theory.~\cite{Baroni} The electron dielectric functions were calculated within the random-phase approximation, where the frequency dependence was retained. $\mathcal{K}^{\rm ph}$ and $\mathcal{Z}^{\rm ph}$ were calculated with the $n${\bf k}-averaged approximate formula [Eq.~(23) in Ref.~\onlinecite{SCDFTII} and Eq.~(40) in Ref.~\onlinecite{Akashi-ph-asym}, respectively], whereas $\mathcal{K}^{\rm el}$ was calculated including the plasmon-induced dynamical screening effect~\cite{Akashi-plasmon-PRL,Akashi-plasmon-JPSJ}. The SCDFT gap equation was solved with the random sampling scheme given in Ref.~\onlinecite{Akashi-MNCl}, with which the sampling error was approximately a few \%. Further details are summarized in Appendix~\ref{sec:detail}. 

\begin{figure}[t!]
 \begin{center}
  \includegraphics[scale=0.9]{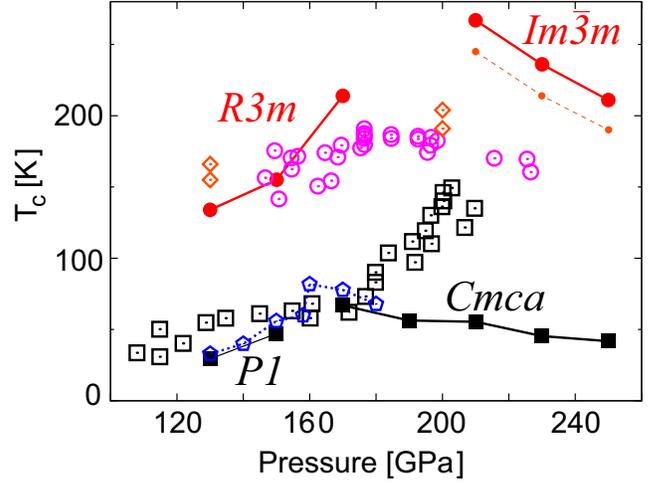}
  \caption{Calculated superconducting transition temperatures for H$_{2}$S (solid square) and H$_{3}$S (solid circle). Experimentally observed values for H$_{2}$S [Fig. 2(a) (open square) and Fig.~2(b) (open circle) of Ref.~\onlinecite{Eremets}] are also plotted together, where different runs are represented by the same symbols. Open pentagon and diamond denote the {\it ab initio} predictions for H$_{2}$S~\cite{Li-Ma-H2S-struct} and H$_{3}$S,~\cite{Duan-H3S-struct} respectively. The small solid circle for the $Im\overline{3}m$-H$_{3}$S phase indicates the calculated result without the contribution of the plasmon mechanism.}
  \label{fig:SHx-Tc-exp}
 \end{center}
\end{figure}

We took particular care for calculating the Eliashberg function 
\begin{eqnarray}
\!\!\!\!\alpha^{2}\!F(\!\omega\!)\!\!=\!\!\frac{1}{N(0)}\!\!\!\sum_{\substack{\nu {\bf q} \\ nn'{\bf k}}}\!\!|g^{n{\bf k}+{\bf q},n'\!{\bf k}}_{\nu{\bf q}}|^{2}
\delta\!(\xi_{n{\bf k}+{\bf q}})\delta\!(\xi_{n'{\bf k}})
\delta\!(\omega\!-\!\omega_{\nu{\bf q}})
\label{eq:a2F}
\end{eqnarray}
employed for $\mathcal{K}^{\rm ph}$ and $\mathcal{Z}^{\rm ph}$. $N(0)$, $g^{n{\bf k}+{\bf q},n'\!{\bf k}}_{\nu{\bf q}}$ and $\omega_{\nu{\bf q}}$ denote the density of states at the Fermi energy, the electron-phonon matrix element and the phonon frequency, respectively.  Since we have found that $\alpha^{2}F(\omega)$ sensitively depends on the smearing scheme and ${\bf k}$- and ${\bf q}$-point density for the integration, we employed a recently developed tetrahedron method with optimized linear interpolation.~\cite{opt-tetra}

We included the plasmon-induced frequency dependence of the screened Coulomb interaction in $\mathcal{K}^{\rm el}$ with the following formula [Eq.~(2) of Ref.~\onlinecite{Akashi-plasmon-PRL}]
\begin{eqnarray}
&&
\mathcal{K}^{\rm el, dyn}_{n{\bf k},n'{\bf k}}
\!=\!
\lim_{\Delta_{n{\bf k}}\rightarrow 0}
\frac{1}{{\rm tanh}[(\beta /2 ) E_{n{\bf k}}]}
\frac{1}{{\rm tanh}[(\beta /2) E_{n'{\bf k}'}]}
\frac{1}{\beta^{2}}
\nonumber \\
&&
\hspace{10pt}\times
\sum_{\omega_{1}\omega_{2}}
F_{n{\bf k}}({\rm i}\omega_{1})
F_{n{\bf k}}({\rm i}\omega_{2})
W_{n{\bf k}n'{\bf k}'}[{\rm i}(\omega_{1}\!\!-\!\!\omega_{2})]
,
\end{eqnarray}
where $W_{n{\bf k}n'{\bf k}'}[{\rm i}(\omega_{1}\!\!-\!\!\omega_{2})]$ is the screened Coulomb interaction and $F_{n{\bf k}}({\rm i}\omega)$$=$
$\frac{1}{{\rm i}\omega\!+\!E_{n{\bf k}}}
\!-\!
\frac{1}{{\rm i}\omega\!-\!E_{n{\bf k}}}
$ denote the electronic anomalous Green's function. In the previous calculations,~\cite{Akashi-plasmon-PRL, Akashi-plasmon-JPSJ} we carried out the Matsubara summations analytically by approximating $W_{n{\bf k}n'{\bf k}'}[{\rm i}(\omega_{1}\!\!-\!\!\omega_{2})]$ with model functions. In the present study, the summation for $\omega_{1}$ was done analytically with variable transformation $\omega_{1}-\omega_{2}\equiv\nu$, whereas the summation for $\nu$ was evaluated numerically with $\sum_{\nu}\sim \frac{1}{T}\int {\rm d}\nu$ without any modeling of $W_{n{\bf k}n'{\bf k}'}[{\rm i}(\omega_{1}\!\!-\!\!\omega_{2})]$, where $T$ is the temperature.\cite{comment-cutoff} 

\section{Results and Discussion}
\label{sec:result-discussion}
Below, we show the calculated values of $T_{\rm c}$ and key factors for the phonon theory: $\lambda$, $\omega_{\rm ln}$, $\mu^{\ast}$ and the isotope effect coefficient $\alpha$. The specific values are summarized in Appendix~\ref{sec:data}.

\begin{figure}[t!]
 \begin{center}
  \includegraphics[scale=0.65]{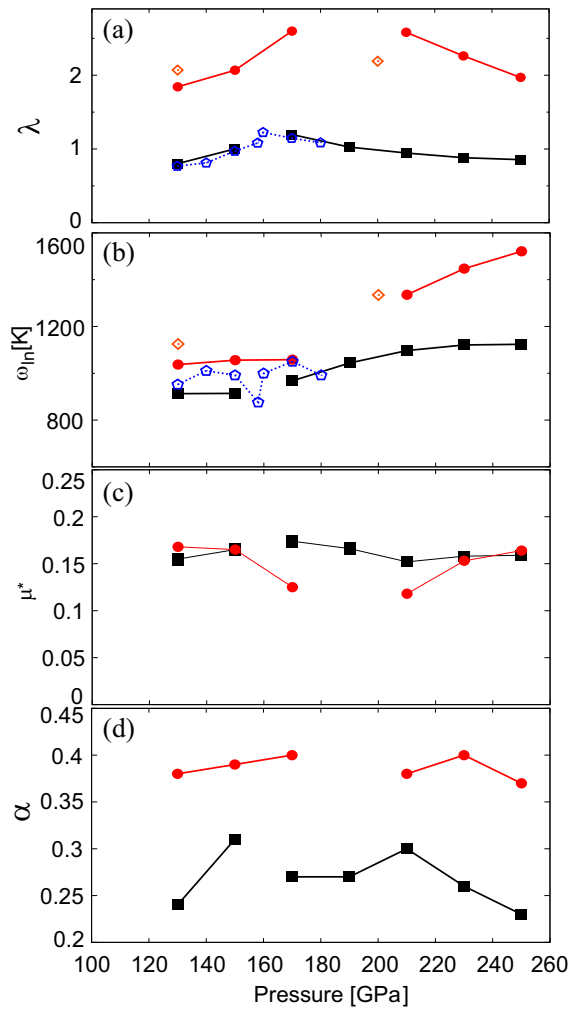}
  \caption{Key factors in the conventional theory for the phonon mechanism calculated from first principles: (a) $\lambda$, (b) $\omega_{\rm ln}$, (c) $\mu^{\ast}$ and (d) $\alpha$. Solid square (circle) denote the values for H$_{2}$S (H$_{3}$S). Open pentagon and diamond represent the preceding {\it ab initio} calculations for H$_{2}$S~\cite{Li-Ma-H2S-struct} and H$_{3}$S,~\cite{Duan-H3S-struct} respectively.}
  \label{fig:SHx-all}
 \end{center}
\end{figure}

In Fig.~\ref{fig:SHx-Tc-exp}, we show the calculated $T_{\rm c}$ with the previously published experimental and first-principles numerical data.~\cite{Eremets, Li-Ma-H2S-struct,Duan-H3S-struct} Drozdov and coworkers~\cite{Eremets} reported two data groups obtained with different experimental conditions, which are indicated by open square and circle, respectively; in this work, we name these groups data 1 and data 2, respectively. The calculated $T_{\rm c}$s for H$_{2}$S (solid square) and H$_{3}$S (solid circle) were $\sim$50~K and $\geq$130~K, respectively. For both H$_{2}$S and H$_{3}$S, the calculated $T_{\rm c}$s show domelike dependence on the pressure. The maximum $T_{\rm c}$s are achieved near the theoretically proposed structural transition points.~\cite{Li-Ma-H2S-struct,Duan-H3S-struct} Our calculated values are as a whole in good agreement with the previous estimates with the McMillan-Allen-Dynes formula (Refs.~\onlinecite{Li-Ma-H2S-struct} and \onlinecite{Duan-H3S-struct}). Notably, for H$_{3}$S, we obtained 267~K at maximum, which is larger by $\sim$60~K from the previous estimate~\cite{Duan-H3S-struct} at 200~GPa. This difference is discussed more specifically later. In the low-pressure regime, the calculated $T_{\rm c}$ for H$_{2}$S (H$_{3}$S) agrees well with data 1 (data 2). In the high-pressure regime, on the other hand, the calculated values are too high or too low compared with the experimental ones. Furthermore, the rapidly increasing feature of data 1 ($\gtrsim$~170~GPa) was not reproduced. We also revisit this point later. Regarding the plasmon effect,~\cite{Akashi-plasmon-PRL,Akashi-plasmon-JPSJ} the enhancement of $T_{\rm c}$ was estimated to be 15--20\% ($\sim$10\%) for H$_{2}$S (H$_{3}$S) (e.g., see small solid circle).

\begin{figure}[b!]
 \begin{center}
  \includegraphics[scale=0.65]{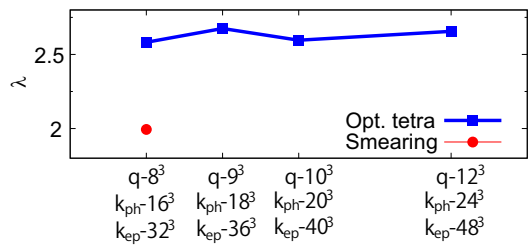}
  \caption{Numerical convergence of $\lambda$ with different schemes for the phonon and $\alpha^{2}F(\omega)$ calculations: Optimized tetrahedron and the 1st-order Hermite-Gaussian smearing with width of 0.030~Ry. ${\bf k}_{\rm ph}$ and ${\bf k}_{\rm ep}$ represent the k-point grids employed for the phonon dynamical matrix and Eliashberg function, respectively. The ${\bf q}$-point summation for ``Smearing" was done with a ${\bf q}$-point grid without offset.}
  \label{fig:lambda-conv}
 \end{center}
\end{figure}

To understand the pressure dependence of the calculated $T_{\rm c}$ in terms of the McMillan-Allen-Dynes formula,~\cite{Allen-Dynes} we show the calculated values of $\lambda=2\int d\omega \frac{\alpha^{2}F(\omega)}{\omega}$ and $\omega_{\rm ln}={\rm exp}[\frac{2}{\lambda}\int d\omega \frac{\alpha^{2}F(\omega){\rm ln}\omega}{\omega}]$ in Figs.~\ref{fig:SHx-all}~(a) and (b), respectively. We see that the pressure dependence of the present {\it ab initio} $T_{\rm c}$s for H$_{2}$S and H$_{3}$S are similar to that of $\lambda$, which indicates that the $T_{\rm c}$s of the present systems are governed by $\lambda$. With this plot for $\lambda$, we see that our tetrahedron method~\cite{opt-tetra} and the previously employed Gaussian smearing scheme~\cite{Duan-H3S-struct,M-P} give different values for $\lambda$, which results in the large difference in $T_{\rm c}$. In fact, by calculating $\lambda$ with the first-order Hermite-Gaussian approximate function $[\delta(\xi)\simeq\frac{1}{\sqrt{\pi} W}[3/2-\left(\xi / W \right)^{2}]{\rm exp}(-\left(\xi / W \right)^{2})$ with $W$$=$0.030~Ry, Ref.~\onlinecite{M-P}], we obtained $\lambda$$=$2.23 and 1.99 for $P=$200~GPa and 210~GPa, respectively, which is consistent with the previous value ($\lambda$$=$$2.19$ for $P$$=$$200$~GPa~\cite{Duan-H3S-struct}). Since the bandwidth of the electronic states is extremely large and complex-shaped electron/hole pockets emerge in this system,~\cite{Duan-H3S-struct} the present tetrahedron-interpolation-based method is expected to be more numerically accurate. We have confirmed the numerical convergence of $\lambda$ as depicted in Fig.~\ref{fig:lambda-conv}. $\omega_{\rm ln}$ monotonically increases as the pressure is increased, which represents the hardening of phonons by compression. This hardening is responsible for the marked difference in $T_{\rm c}$s for $R3m$-H$_{3}$S and $Im\overline{3}m$-H$_{3}$S. For higher pressure regime, however, the hardening is dominated over by the decrease of $\lambda$ and therefore $T_{\rm c}$ decreases.~\cite{Pickett-preprint}

We determined optimum values for $\mu^{\ast}$ so that the $T_{\rm c}$s calculated with the SCDFT gap equation can be reproduced with the extended McMillan formula.~\cite{Allen-Dynes} For H$_{2}$S, the optimum values were 0.15--0.17 for all the pressures. For the pressure range 170--210~GPa, we observed decrease of the optimum values for H$_{3}$S. Probably this is originating from a fact that $T_{\rm c}$ calculated by the present SCDFT sometimes deviates slightly from that by the Eliashberg equation.~\cite{SCDFTI} Detailed investigations on this point are left for future studies.

Using the calculated $T_{\rm c}$s for H$_{x}$S and D$_{x}$S, we also calculated the isotope effect coefficient $\alpha=-[{\rm ln}T_{\rm c}^{{\rm D}_{x}{\rm S}}-{\rm ln}T_{\rm c}^{{\rm H}_{x}{\rm S}}]/[{{\rm ln}M_{\rm D}-{\rm ln}M_{\rm H}}]$, where $T_{\rm c}^{{\rm H}_{x}{\rm S}}$~($T_{\rm c}^{{\rm D}_{x}{\rm S}}$) is the transition temperature of the hydride (deuteride) compound and $M_{\rm H}$ ($M_{\rm D}$)  is the atomic mass of hydrogen (deuterium), respectively. The values ranges between 0.23 and 0.31 (0.38 and 0.42) for H$_{2}$S (H$_{3}$S). These values are smaller than the BCS value ($\alpha$$\sim$$0.5$), which indicate the correction due to the retardation effect.

Here we compare our calculated and experimentally observed values of $T_{\rm c}$. First, the experimentally observed $T_{\rm c}$s in the low-pressure regime were quantitatively reproduced by assuming the emergence of single structural phases of $P1$-H$_{2}$S and $R3m$-H$_{3}$S for data 1 and 2, respectively. This strongly suggests that these two phases are dominant in the experimental situations for $P$$\lesssim$150~GPa. It is even conceivable that the high-pressure values of data 2 corresponds to $R3m$-H$_{3}$S. The agreement of the calculated and experimentally observed $T_{\rm c}$s for higher pressures were, on the other hand, not as perfect as those for the previously studied conventional superconductors.~\cite{SCDFTII,Floris-MgB2,Sanna-CaC6,Profeta-Li,Floris-Pb,Akashi-plasmon-PRL} Note that we assumed that the sample is homogeneous and does not decompose into H$_{x}$S and S for all the pressure range, though it has not been confirmed experimentally. Our calculated $T_{\rm c}$ for $Im\overline{3}m$-H$_{3}$S suggests that maximum $T_{\rm c}$ can be increased to, possibly, a higher value in the pure $Im\overline{3}m$-H$_{3}$S phase. 

Very recently, there has been an independent report on an {\it ab initio} $T_{\rm c}$ calculation for $Im\overline{3}m$-H$_{3}$S using the SCDFT~\cite{Gross-preprint} with a condition different from ours.~\cite{comment-condition} They concluded that the experimentally observed high $T_{\rm c}$ can be explained with $Im\overline{3}m$-H$_{3}$S, whereas we rather propose a relevance of $R3m$-H$_{3}$S in the experimental situation.

Finally, we move on to $\alpha$. The calculated values were far smaller than the experimentally observed $\alpha$$\sim$$1.0$. Based on a hypothesis of inhomogeneity, let us give a possible explanation of the experimental large $\alpha$ within the present framework. As suggested by Hirsch and Marsiglio,~\cite{Hirsch-Marsiglio-preprint} when inhomogeneity of the system is substantial, the experimentally observed $T_{\rm c}$ should somehow deviate. For example, suppose we estimate $\alpha$ with $\alpha=-[{\rm ln}T_{\rm c}^{{\rm D}_{2}{\rm S}}-{\rm ln}T_{\rm c}^{{\rm H}_{3}{\rm S}}]/[{{\rm ln}M_{\rm D}-{\rm ln}M_{\rm H}}]$; we then get $\alpha$$\gtrsim 2.0$ for the whole pressure range. Such a situation is possible because the enthalpy difference between H$_{2}$S and $\frac{2}{3}$H$_{3}$S+$\frac{1}{3}$S is of order of the phonon frequency~\cite{Mazin-preprint}: Substitution of D for H substantially modulate the contribution of the zero-point oscillation to the total enthalpy and it should change the relative stability of the competing phases. 

We thus suggest that the H$_{3}$S phases have a key role for understanding the reported experimental results~\cite{Eremets} and realizing higher $T_{\rm c}$. To validate/invalidate this, measurements with different chemical composition (e.g., H:S=3:1) and compression at higher temperatures might be helpful. When measuring the isotope effect, the difference in the structural relaxation speed of hydrides and deuterides should also be taken into account.

\begin{table*}[t!]
\caption[t]
{Detailed settings for the calculations. Subscript ``1" for {\bf q} points denotes the mesh with displacement by half a grid step.}
\begin{center}
\label{tab:conditions}
\tabcolsep = 1mm
\begin{tabular}{|l |c|c|c|c|c|} \hline
 &&$P1$-H$_{2}$S & $Cmca$-H$_{2}$S & $R3m$-H$_{3}$S & $Im\overline{3}m$ H$_{3}$S \\ \hline
charge density &{\bf k} & (12 12 8) & (12 12 4)& (16 16 16)&(16 16 16) \\ \cline{2-6}
                      &interpol. &\multicolumn{4}{c|}{1$^{\rm st}$ order Hermine Gaussian~\cite{M-P} with width=0.030Ry} \\ \hline
dynamical matrix &{\bf k} & (12 12 8) & (12 12 4)& (16 16 16)&(16 16 16) \\ \cline{2-6}
                        &{\bf q} & (6 6 4)$_{1}$ & (6 6 2)$_{1}$& (8 8 8)$_{1}$ &(8 8 8)$_{1}$ \\ \cline{2-6}
                        &interpol. &\multicolumn{4}{c|}{Optimized tetrahedron~\cite{opt-tetra}} \\ \hline
electron-phonon &{\bf k}$^{\dagger}$ & (12 12 8) & (24 24 8)& (32 32 32)&(32 32 32) \\ \cline{2-6}
                        &interpol. &\multicolumn{4}{c|}{Optimized tetrahedron~\cite{opt-tetra}} \\ \hline
dielectric function &{\bf k}  for bands crossing $E_{\rm F}^{\dagger\dagger}$ & (18 18 12) & (18 18 6)& (18 18 18) &(18 18 18) \\ \cline{2-6}
                           &{\bf k}  for other bands& (6 6 4) & (6 6 2)& (6 6 6) &(6 6 6) \\ \cline{2-6}
                           &{\bf q}  & (6 6 4) & (6 6 2)& (6 6 6) &(6 6 6) \\ \cline{2-6}
                           &unoccupied band num.& $\sim$60&$\sim$100&$\sim$30& $\sim$30 \\ \cline{2-6}
                           &interpol. &\multicolumn{4}{c|}{Tetrahedron with the Rath-Freeman treatment~\cite{Rath-Freeman}} \\ \hline
SCDFT gap function & unoccupied band num. &25&45&19&19 \\ \cline{2-6} 
           &{\bf k} for $\mathcal{K}^{\rm el}$ & (6 6 4) & (6 6 2)& (6 6 6) &(6 6 6) \\ \cline{2-6} 
           & $N_{\rm s}$ for bands crossing $E_{\rm F}$ &4500&3000&6000&6000 \\ \cline{2-6} 
           & $N_{\rm s}$ for other bands &150&100&200&200 \\ \cline{2-6}
           & Sampling error in $T_{\rm c}$ &$\sim$9\%&$\sim$6\%&$\sim$5\%&$\sim$5\% \\ \hline 
           \multicolumn{6}{l}{$^\dagger$ Electron energy eigenvalues and eigenfunctions were calculated on these auxiliary grid points.} \\
           \multicolumn{6}{l}{$^{\dagger\dagger}$ Electron energy eigenvalues were calculated on these auxiliary grid points.} \\
\end{tabular}
\end{center}
\end{table*}

\section{Summary}
\label{sec:summary}
In this study, we have performed a present state-of-the-art {\it ab initio} calculation for the superconductivity in H$_{2}$S and H$_{3}$S assuming the conventional phonon mechanism, where the effect of the electron-electron Coulomb repulsion was non-empirically treated. For the pressures $\lesssim$~150~GPa, the calculated $T_{\rm c}$s for $P1$-H$_{2}$S and $R3m$-H$_{3}$S agree well with the experimental $T_{\rm c}$s observed with different compressing and cooling conditions, respectively. This strengthens the scenario that H$_{3}$S is superconducting when the high $T_{\rm c}$ is observed.~\cite{Mazin-preprint,Duan-preprint} For the high-pressure phase of $Im\overline{3}m$-H$_{3}$S, we have predicted $T_{\rm c}$ higher than the experimentally observed maximum of 190~K and the values calculated for $R3m$-H$_{3}$S, which amounts to 267~K. This suggests that higher $T_{\rm c}$ can be achieved by isolating the single $Im\overline{3}m$-H$_{3}$S phase. Although we have ignored several possible effects in the present systems (e.g., zero-point oscillation of hydrogen atoms, anharmonic phonons, etc.), the present result can be a key step for further theoretical and experimental investigations on the superconducting sulfur hydrides. Examinations of anharmonic lattice-dynamical effects, which has been neglected with the present methodology, are under way.

\subsection*{Note added}
After the submission of the present work, there has been a publication demonstrating that the anharmonic effect reduces $T_{\rm c}$ by about 20~\% in $Im\overline{3}m$-H$_{3}$S.~\cite{Errea-anharmonic} Nevertheless, the present indications of the relevance of the $R3m$ phase and possible higher $T_{\rm c}$ are still valid since the increase of $T_{\rm c}$ by the plasmon mechanism will compensate the anharmonic effect.

\begin{acknowledgments}
This work is partially supported by Tokodai Institute for Element Strategy (TIES). Y.N. is supported by Grant-in-Aid for JSPS Fellows (Grant No. 12J08652) from Japan Society for the Promotion of Science (JSPS), Japan. We are indebted to Yinwei Li for sharing the pseudopotentials and optimized structural parameters. We also thank Terumasa Tadano for fruitful discussions.
\end{acknowledgments}

\begin{appendix}
\section{Computational detail}
\label{sec:detail}
For the electronic and lattice-dynamical calculations, we used the pseudopotentials for S and H atoms implemented with the Troullier-Martin scheme,~\cite{TM-FHI} which are the same as those used in Ref.~\onlinecite{Li-Ma-H2S-struct}. The plane-wave energy cutoff was set to 80~Ry, whereas the auxiliary cutoff for the dielectric function was 12.8~Ry. Conditions for the calculations of the charge density, dynamical matrix, electron-phonon coupling, dielectric function and gap function are detailed in Table~\ref{tab:conditions}.

\newpage

\section{Numerical data of the calculated values of $T_{\rm c}$, $\lambda$, $\omega_{\rm ln}$, $\mu^{\ast}$ and $\alpha$}
\label{sec:data}
Tables~\ref{tab:Tc}--\ref{tab:alpha} lists the calculated values for $T_{\rm c}$, $\lambda$, $\omega_{\rm ln}$, $\mu$ and $\alpha$.

\begin{table}[h]
\caption[t]
{Superconducting transition temperature $T_{\rm c}$ [K].}
\begin{center}
\label{tab:Tc}
\tabcolsep = 1mm
\begin{tabular}{|l |c|c|c|c|c|c|c|} \hline
 $P$ [GPa]&130&150&170&190&210&230&250 \\ \hline
H$_{2}$S & 29.4& 47.1 &66.9&56.3&55.4&45.4&41.8 \\ \hline
D$_{2}$S & 25.0& 38.2 &55.7&46.7&45.0&37.9&35.7 \\ \hline
H$_{3}$S & 134&155&214&. . .&267&236&211 \\ \hline
D$_{3}$S & 103&119&163&. . .&206&180&164 \\
 \hline
\end{tabular}
\end{center}
\end{table}

\begin{table}[h]
\caption[t]
{Electron-phonon coupling coefficient $\lambda$.}
\begin{center}
\label{tab:lambda}
\tabcolsep = 1mm
\begin{tabular}{|l |c|c|c|c|c|c|c|} \hline
 $P$ [GPa]&130&150&170&190&210&230&250 \\ \hline
H$_{2}$S &0.801 &1.001&1.196&1.026&0.945&0.882&0.855 \\ \hline
H$_{3}$S &1.843 &2.067&2.599&. . .&2.582&2.263&1.970 \\
 \hline
\end{tabular}
\end{center}
\end{table}

\begin{table}[h]
\caption[t]
{Logarithmic moment of the Eliashberg function $\omega_{\rm ln}$ [K].}
\begin{center}
\label{tab:omegaln}
\tabcolsep = 1mm
\begin{tabular}{|l |c|c|c|c|c|c|c|} \hline
 $P$ [GPa]&130&150&170&190&210&230&250 \\ \hline
H$_{2}$S &913 &914&968&1044&1097&1121&1124 \\ \hline
H$_{3}$S &1037 &1056&1058&. . .&1336&1447&1521 \\
 \hline
\end{tabular}
\end{center}
\end{table}

\begin{table}[h]
\caption[t]
{Renormalized electron-electron Coulomb parameter $\mu^{\ast}$ estimated from the $T_{\rm c}$ calculated with the SCDFT gap equation.}
\begin{center}
\label{tab:mustar}
\tabcolsep = 1mm
\begin{tabular}{|l |c|c|c|c|c|c|c|} \hline
 $P$ [GPa]&130&150&170&190&210&230&250 \\ \hline
H$_{2}$S & 0.155&0.165&0.174&0.166&0.152&0.158&0.159 \\ \hline
H$_{3}$S & 0.168&0.165&0.125&. . .&0.118&0.153&0.164 \\
 \hline
\end{tabular}
\end{center}
\end{table}
\newpage
\begin{table}[h]
\caption[t]
{Isotope-effect coefficient $\alpha$.}
\begin{center}
\label{tab:alpha}
\tabcolsep = 1mm
\begin{tabular}{|l |c|c|c|c|c|c|c|} \hline
 $P$ [GPa]&130&150&170&190&210&230&250 \\ \hline
H$_{2}$S & 0.24&0.31&0.27&0.27&0.30&0.26&0.23 \\ \hline
H$_{3}$S & 0.38&0.39&0.40&. . .&0.38&0.40&0.37 \\
 \hline
\end{tabular}
\end{center}
\end{table}
\end{appendix}


\begin{thebibliography}{999}
\bibitem{BCS} J. Bardeen, L. N. Cooper, and J. R. Schrieffer, Phys. Rev. \textbf{108}, 1175 (1957).
\bibitem{fullerene} A. F. Hebard, M. J. Rosseinsky, R. C. Haddon, D. W. Murphy, S. H. Glarum, T. T.
M. Palstra, A. P. Ramirez, and A. R. Kortan, Nature (London) \textbf{350}, 600 (1991).
\bibitem{MgB2} J. Nagamatsu, N. Nakagawa, T. Muranaka, Y. Zenitani, and J. Akimitsu, Nature
(London) \textbf{410}, 63 (2001).
\bibitem{Li-pressure-Shimizu} K. Shimizu, H. Ishikawa, D. Takao, T. Yagi, and K. Amaya, Nature (London) 419,
597 (2002).
\bibitem{Li-pressure-Struzhkin} V.V. Struzhkin, M. I. Eremets, W. Gan, H. K. Mao, and R. J. Hemley, Science 298,
1213 (2002).
\bibitem{B-diamond-Ekimov} E. A. Ekimov, V. A. Sidorov, E. D. Bauer, N. N. Mel'nik, N. J. Curro, J. D. Thompson, and S. M. Stishov, Nature (London) \textbf{428}, 542 (2004).
\bibitem{B-diamond-Takano} H. Okazaki, T. Wakita, T. Muro, T. Nakamura, Y. Muraoka, T. Yokoya, S. Kurihara, H. Kawarada, T. Oguchi, and Y. Takano, arXiv:1411.7752. 
\bibitem{Ashcroft-H} N. W. Ashcroft, Phys. Rev. Lett. \textbf{21}, 1748 (1968).
\bibitem{Cudazzo-hydrogen} P. Cudazzo, G. Profeta, A. Sanna, A. Floris, A. Continenza, S. Massidda, and E. K. U. Gross, Phys. Rev. Lett. \textbf{100}, 257001 (2008).
\bibitem{Ashcroft-hydrides} N. W. Ashcroft, Phys. Rev. Lett. \textbf{92}, 187002 (2004).
\bibitem{Feng-Ashcroft-SiH4} J. Feng, W. Grochala, T. Jaro\'n, R. Hoffmann, A. Bergara, and N. W. Ashcroft, Phys. Rev. Lett. \textbf{96}, 017006 (2006).
\bibitem{Tse-SiH4-SC} Y. Yao, J. S. Tse, Y. Ma, K. Tanaka, Europhys. Lett. \textbf{78}, 37003 (2007).
\bibitem{Eremets-Tse-SiH4-exp2008} M. I. Eremets, I. A. Trojan, S. A. Medvedev, J. S. Tse, and Y. Yao, Science \textbf{319}, 1506 (2008).
\bibitem{Degtyareva-PtH-exp2009} O. Degtyareva, J. E. Proctor, C. L. Guillaume, E. Gregoryanz, and M. Hanfland, Solid State Commun. \textbf{149}, 1583 (2009).
\bibitem{Jin-Ma-Cui-Si2H6-PNAS2010} X. Jin, X. Meng, Z. He, Y. Ma, B. Liu, T. Cui, G. Zou, and H. Mao, Proc. Natl. Acad. Sci. USA \textbf{107}, 9969 (2010).
\bibitem{Li-Gao-SiH4H22-PNAS2010} Y. Li, G. Gao, Y. Xie, Y. Ma, T. Cui, and G. Zou, Proc. Natl. Acad. Sci. USA \textbf{107}, 15708 (2010).
\bibitem{Jose-Marques-Goedecker-Si2H6-PRL2012} J. A. Flores-Livas, M. Amsler, T. J. Lenosky, L. Lehtovaara, S. Botti, M. A. L. Marques, and S. Goedecker, Phys. Rev. Lett. \textbf{108}, 117004 (2012).
\bibitem{Scheler-Degtyareva-PtH-PRB2011} T. Scheler, O. Degtyareva, M. Marqu\'es, C. L. Guillaume, J. E. Proctor, S. Evans, and E. Gregoryanz, Phys. Rev. B \textbf{83}, 214106 (2011).
\bibitem{Zhou-Oganov-PtH-PRB2011} X. F. Zhou, A. R. Oganov, X. Dong, L. Zhang, Y. Tian, and H. T. Wang, Phys. Rev. B \textbf{84}, 054543(2011).
\bibitem{Kim-Pickard-Needs-PtH-PRL2011} D. Y. Kim, R. H. Scheicher, C. J. Pickard, R. J. Needs, and R. Ahuja, Phys. Rev. Lett. \textbf{107}, 117002 (2011). 
\bibitem{Tse-SnH4-PRL2007} J. S. Tse, Y. Yao, and K. Tanaka, Phys. Rev. Lett. \textbf{98}, 117004 (2007).
\bibitem{Gao-Oganov-SnH4-PNAS2010} G. Gao, A. R. Oganov, P. Li, Z. Li, H. Wang, T. Cui, Y. Ma, A. Bergara, A. O. Lyakhov, T. Iitaka, and G. Zou, Proc. Natl. Acad. Sci. USA \textbf{107}, 1317 (2010)
\bibitem{Goncharenko-Erements-Tse-AlH3-PRL2008} I. Goncharenko, M. I. Eremets, M. Hanfland, J. S. Tse, M. Amboage, Y. Yao, and I. A. Trojan, Phys. Rev. Lett. \textbf{100}, 045504 (2008).
\bibitem{Gao-Oganov-GeH4-PRL2008} G. Gao, A. R. Oganov, A. Bergara, M. Martinez-Canales, T. Cui, T. Iitaka, Y. Ma, and G. Zou, Phys. Rev. Lett. \textbf{101}, 107002 (2008).
\bibitem{Szczesniak-GeH4-SSComm2013} R. Szcz\c{e}\'sniak, A. P. Durajski, and D. Szcz\c{e}\'sniak, Solid State Commun. \textbf{165}, 39 (2013).
\bibitem{Gao-Bergara-GaH3-PRB2011} G. Gao, H. Wang, A. Bergara, Y. Li, G. Liu, and Y. Ma, Phys. Rev. B \textbf{84}, 064118 (2011).
\bibitem{Szczesniak-GaH3-SSTech2014} R. Szcz\c{e}\'sniak and A. P. Durajski, Supercond. Sci. Technol. \textbf{27}, 015003 (2014).
\bibitem{Tse-CaH-PNAS} H. Wang, J. S. Tse, K. Tanaka, T. Iitaka, and Y. Ma, Proc. Natl. Acad. Sci. USA \textbf{109}, 6463 (2012).
\bibitem{Lonie-Zurek-MgH-PRB2013} D. C. Lonie, J. Hooper, B. Altintas, and E. Zurek, Phys. Rev. B \textbf{87}, 054107 (2013).
\bibitem{Wang-Yao-Ma-BeH2-JChemPhys2014} Z. Wang, Y. Yao, L. Zhu, H. Liu, T. Iitaka, H. Wang, and Y. Ma, J. Chem. Phys. \textbf{140}, 124707 (2014).
\bibitem{Zhou-Ma-Cui-KH6-PRB2012} D. Zhou, X. Jin, X. Meng, G. Bao, Y. Ma, B. Liu, and T. Cui, Phys. Rev. B \textbf{86}, 014118 (2012).
\bibitem{Kim-Ahuja-trend-PNAS2010} D. Y. Kim. R. H. Scheicher, H. Mao, T. W. Kang, and R. Ahuja, Proc. Natl. Acad. Sci. USA \textbf{107}, 2793 (2010).
\bibitem{Gao-Ashcroft-Bergara-NbH-PRB2013} G. Gao, R. Hoffmann, N. W. Ashcroft, H. Liu, A. Bergara, and Y. Ma, Phys.Rev. B \textbf{88}, 184104 (2013).
\bibitem{Li-Ma-H2S-struct} Y. Li, J. Hao, H. Liu, Y. Li, and Y. Ma, J. Chem. Phys. \textbf{140}, 174712 (2014).
\bibitem{Duan-H3S-struct} D. Duan, Y. Liu, F. Tian, D. Li, X. Huang, Z. Zhao, H. Yu, B. Liu, W. Tian, and T. Cui, Sci. Reports \textbf{4}, 6968 (2014). 
\bibitem{Eremets} A. P. Drozdov, M. I. Eremets, and I. A. Troyan, arXiv:1412.0460.
\bibitem{Hirsch-Marsiglio-preprint} J. E. Hirsch and F. Marsiglio, Physica C \textbf{511}, 45 (2015).
\bibitem{Durajski-Li-preprint} A. P. Durajski, R. Szcz\c{e}\'sniak, and Y. Li, Physica C \textbf{515}, 1 (2015).
\bibitem{Mazin-preprint} N. Bernstein, C. S. Hellberg, M. D. Johannes, I. I. Mazin, and M. J. Mehl, Phys. Rev. B \textbf{91}, 060511(R) (2015).
\bibitem{Duan-preprint} D. Duan, X. Huang, F. Tian, D. Li, H. Yu, Y. Liu, Y. Ma, B. Liu, and T. Cui, Phys. Rev. B \textbf{91}, 180502(R) (2015).
\bibitem{Pickett-preprint} D. A. Papaconstantopoulos, B. M. Klein, M. J. Mehl, and W. E. Pickett, Phys. Rev. B \textbf{91}, 184511 (2015).
\bibitem{Migdal} A. B. Migdal, Sov. Phys. JETP \textbf{7}, 996 (1958).
\bibitem{Allen-Dynes} P. B. Allen and R. C. Dynes, Phys. Rev. B \textbf{12}, 905 (1975).
\bibitem{Morel-Anderson} P. Morel and P. W. Anderson, Phys. Rev. \textbf{125}, 1263 (1962).
\bibitem{Garland-isotope} J. W. Garland, Phys. Rev. Lett. \textbf{11}, 114 (1963)
\bibitem{Carbotte-review} J. P. Carbotte, Rev. Mod. Phys. \textbf{62}, 1027 (1990).
\bibitem{ME} G. M. Eliashberg, Sov. Phys. JETP \textbf{11}, 696 (1960); D. J. Scalapino, in {\it Superconductivity} edited by R. D. Parks, (Marcel Dekker, New York, 1969) VOLUME 1; J. R. Schrieffer,{\it Theory of superconductivity; Revised Printing},  (Westview Press, Colorado, 1971).
\bibitem{SCDFTI} M. L\"uders, M. A. L. Marques, N. N. Lathiotakis, A. Floris, G. Profeta, L. Fast, A. Continenza, S. Massidda, and E. K. U. Gross, Phys. Rev. B \textbf{72}, 024545 (2005).
\bibitem{SCDFTII} M. A. L. Marques, M. L\"uders, N. N. Lathiotakis, G. Profeta, A. Floris, L. Fast, A. Continenza, E. K. U. Gross, and S. Massidda, Phys. Rev. B \textbf{72}, 024546 (2005).
\bibitem{Akashi-plasmon-PRL} R. Akashi and R. Arita, Phys. Rev. Lett. \textbf{111}, 057006 (2013).
\bibitem{Akashi-plasmon-JPSJ} R. Akashi and R. Arita, J. Phys. Soc. Jpn. \textbf{83}, 061016 (2014).
\bibitem{Takada-plasmon-JPSJ} Y. Takada, J. Phys. Soc. Jpn. \textbf{45}, 786 (1978).
\bibitem{GGAPBE} J. P. Perdew, K. Burke, and M. Ernzerhof, Phys. Rev. Lett. \textbf{77}, 3865 (1996).
\bibitem{QE} P. Giannozzi, S. Baroni, N. Bonini, M. Calandra, R. Car,
C. Cavazzoni, D. Ceresoli, G. L. Chiarotti, M. Cococcioni,
I. Dabo, A. Dal Corso, S. Fabris, G. Fratesi, S.
de Gironcoli, R. Gebauer, U. Gerstmann, C. Gougoussis,
A. Kokalj, M. Lazzeri, L. Martin-Samos, N. Marzari,
F. Mauri, R. Mazzarello, S. Paolini, A. Pasquarello, L.
Paulatto, C. Sbraccia, S. Scandolo, G. Sclauzero, A. P.
Seitsonen, A. Smogunov, P. Umari, and R. M. Wentzcovitch,
J. Phys.: Condens. Matter 21, 395502 (2009);
http://www.quantum-espresso.org/
\bibitem{Baroni}  S. Baroni, S. de Gironcoli, A. Dal Corso, and P. Giannozzi, Rev. Mod. Phys. \textbf{73}, 515(2001).
\bibitem{Akashi-ph-asym} R. Akashi and R. Arita, Phys. Rev. B \textbf{88}, 014514 (2013).
\bibitem{Akashi-MNCl} R. Akashi, K. Nakamura, R. Arita, and M. Imada, Phys. Rev. B \textbf{86}, 054513 (2012).
\bibitem{opt-tetra} M. Kawamura, Y. Gohda, and S. Tsuneyuki, Phys. Rev. B \textbf{89}, 094515 (2014).
\bibitem{comment-cutoff} We employed the approximation $\int {\rm d}\nu$$\simeq$$2\int^{\nu_{\rm max}}_{0} {\rm d}\nu$ for the numerical $\nu$-integral with $\nu_{\rm max}$$=$70~eV. The high-frequency contribution $\int_{\nu_{\rm max}}^{\infty} {\rm d}\nu$ was estimated analytically with the approximation $W_{n{\bf k}n'{\bf k}'}({\rm i}\nu)$$\simeq$$W_{n{\bf k}n'{\bf k}'}({\rm i}\nu_{\rm max})$.
\bibitem{M-P} M. Methfessel and A. T. Paxton, Phys. Rev. B \textbf{40}, 3616 (1989).
\bibitem{Floris-Pb} A. Floris, A. Sanna, S. Massidda, and E. K. U. Gross, Phys. Rev. B \textbf{75}, 054508 (2007).
\bibitem{Floris-MgB2} A. Floris, G. Profeta, N. N. Lathiotakis, M. L\"uders, M. A. L. Marques, C. Franchini, E. K. U. Gross, A. Continenza, and S. Massidda, Phys. Rev. Lett. \textbf{94}, 037004 (2005); A. Floris, A. Sanna, M. L\"uders, G. Profeta, N. N. Lathiotakis, M. A. L. Marques, C. Franchini, E. K. U. Gross, A. Continenza, and S. Massidda, Physica C \textbf{456}, 45 (2007).
\bibitem{Sanna-CaC6} A. Sanna, G. Profeta, A. Floris, A. Marini, E. K. U. Gross, and S. Massidda, Phys. Rev. B \textbf{75}, 020511(R) (2007).
\bibitem{Profeta-Li} G. Profeta, C. Franchini, N. N. Lathiotakis, A. Floris, A. Sanna, M. A. L. Marques, M. L\"uders, S. Massidda, E. K. U. Gross, and A. Continenza, Phys. Rev. Lett. \textbf{96}, 047003 (2006).
\bibitem{Gross-preprint} J. A. Flores-Livas, A. Sanna, and E. K. U. Gross, arXiv:1501.06336.
\bibitem{comment-condition} In Ref.~\onlinecite{Gross-preprint} (i) they used modified forms for the phonon part of the SCDFT kernels, though the detail of the modification has been unpublished yet~(See Ref.~56 in Ref.~\onlinecite{Gross-preprint}), (ii) they did not include the plasmon-induced dynamical effect, and (iii) their calculation was based on the local-density approximation,~\cite{Ceperley-Alder, PZ81} which yielded a $R3m$-$Im\overline{3}m$ structural transition point slightly different from that with the generalized-gradient approximation.~\cite{Duan-H3S-struct}  
\bibitem{Ceperley-Alder} D. M. Ceperley and B. J. Alder, Phys. Rev. Lett. \textbf{45}, 566 (1980).
\bibitem{PZ81} J. P. Perdew and A. Zunger, Phys. Rev. B \textbf{23}, 5048 (1981).
\bibitem{Errea-anharmonic} I. Errea, M. Calandra, C. J. Pickard, J. Nelson, R. J. Needs, Y. Li, H. Liu, Y. Zhang, Y. Ma, F. Mauri, Phys. Rev. Lett. \textbf{114}, 157004 (2015).
\bibitem{TM-FHI} N. Troullier and J. L. Martins, Phys. Rev. B \textbf{43}, 1993 (1991); http://www.abinit.org/downloads/psp-links/psp-links/gga\_fhi .
\bibitem{Rath-Freeman} J. Rath and A. J. Freeman, Phys. Rev. B \textbf{11}, 2109 (1975).
\end{thebibliography}
\end{document}